\documentclass[11pt]{amsart}
\usepackage[T1]{fontenc}
\usepackage[letterpaper,margin=1in]{geometry}
\usepackage{amsmath,amssymb,booktabs,hyperref}
\newtheorem{theorem}{Theorem}[section]
\newtheorem{lemma}[theorem]{Lemma}
\newtheorem{prop}[theorem]{Proposition}
\newtheorem{exam}[theorem]{Example}
\newtheorem{cor}[theorem]{Corollary}

\newtheorem{question}[theorem]{Question}

\hypersetup{hidelinks}
\usepackage{tikz}
\usetikzlibrary{automata,positioning}
\newcommand{\hide}[1]{%
  \smash{\rlap{%
    \parbox[t]{0.8\linewidth}{%
      \color{white}\fontsize{5}{6}\selectfont #1%
    }%
  }}%
}

\title[Boxes in Infinite Words]{Boxes in Infinite Words: Complexity, Automata, and Point Modules}

\author{Jason Bell}
\thanks{The author was supported by a
Discovery Grant from the Natural Sciences and
Engineering Research Council of Canada (NSERC)}
\address{Department of Pure Mathematics,
University of Waterloo,
Waterloo, Ontario, Canada N2L 3G1}
\email{jpbell@uwaterloo.ca}
\begin{document}

\begin{abstract}
Motivated by the geometry of point modules over
monomial algebras, we introduce a box-counting
function associated to a right-infinite word over
a finite alphabet. An extendable box is a family
of finite words obtained by choosing letters
independently at each position, subject to the
condition that all choices admit a common infinite
right continuation in the associated subshift.
We prove several results for these functions for words of linearly bounded complexity.
For $k$-automatic words, we show using techniques from first-order logic that the total
box-counting function is $k$-regular. Finally, we apply our results
to prove eventual affine growth of the number
of positive-dimensional irreducible components
in finite projections of point-module spaces
associated to semiprime monomial algebras of
at most quadratic growth.
\end{abstract}

\maketitle
\section{Introduction}

Let ${\bf w}=w_0w_1w_2\cdots$ be a right-infinite word over a finite
alphabet $\Sigma$. A fundamental invariant of ${\bf w}$ is its
\emph{subword complexity function}
\[
p_{\bf w}(N)=
\#\{u\in\Sigma^N:u\text{ occurs as a consecutive subword of }{\bf w}\}.
\]
Thus $p_{\bf w}(N)$ counts the distinct length-$N$ subwords of
${\bf w}$. Subword complexity has been extensively studied in
symbolic dynamics and the theory of automatic sequences (see, for example, \cite{ASbook,Brlek,Cass,CassaigneNicolas,Ferenczi,GSS}). 

There is a natural dynamical system associated to ${\bf w}$. 
If we endow $\Sigma$ with the discrete topology then $\Sigma^{\mathbb N}$, the space of right-infinite words over $\Sigma$, is a compact topological space when given the product topology.
We have a left shift map $\sigma$ on $\Sigma^{\mathbb N}$, which sends an infinite word $z_0z_1z_2\cdots $ to the word $z_1z_2\cdots$ obtained by removal of the first letter.  We can then create a closed shift-invariant set
\[
X_{\bf w}=\overline{\{\sigma^n({\bf w}):n\ge0\}},
\]
which is the closure of the orbit of ${\bf w}$ under the left shift $\sigma$. This is a one-sided
subshift with the property that the (finite) subwords of an element of $X_{\bf w}$ are finite subwords
of ${\bf w}$. It is therefore natural to view the complexity function $p_{\bf w}(N)$ as measuring the 
number of distinct finite patterns in an associated symbolic dynamical system.

In this paper we study a different counting problem: how many
maximal families of finite patterns $B$ are there with the property that all elements of $B$ share a common right-infinite extension ${\bf z}$ such that $B{\bf z}\subseteq X_{\bf w}$?
Our motivation for this counting problem comes from noncommutative projective geometry. To
an infinite word ${\bf w}$ one can associate a graded monomial
algebra $A_{\bf w}$ whose degree-$N$ homogeneous component has
dimension $p_{\bf w}(N)$. In earlier work of the author and
Greenfeld \cite{BG}, point modules over such algebras were
described concretely in terms of sequences of such finite patterns.
We give further background and details in \S\ref{sec:algebraic-background}. One can understand point modules in terms of their successive truncations and thus the collection of point modules takes the structure of proalgebraic variety (see \cite{BG} for details).   
For a monomial algebra $A_{\bf w}$, the space of truncations of point modules are
finite unions of finite products of projective spaces, and their
irreducible components corresponding to certain associated families
of finite words which we call boxes. These are precisely the families we wish to
count in this paper, and although the original motivation for this problem is geometric, the problems we consider can be
formulated entirely in terms of words and subshifts.

Given nonempty subsets $S_0,\ldots,S_{N-1}\subseteq\Sigma$, put
\[
B=S_0S_1\cdots S_{N-1}
 =\{a_0a_1\cdots a_{N-1}:a_i\in S_i\text{ for all }i\}.
\]
We call $B$ an $N$-\emph{box}, and define its \emph{dimension}
and \emph{size} by
\[
\dim B=\sum_{i=0}^{N-1}(|S_i|-1),
\qquad |B|=\prod_{i=0}^{N-1}|S_i|.
\]
A position $i$ is \emph{branching} if $|S_i|>1$, so a box has
positive dimension precisely when it has a branching position.
For example, over $\Sigma=\{0,1\}$ the box
\[
\{0,1\}\{0\}\{0,1\}=\{000,001,100,101\}
\]
has dimension $2$ and size $4$. The dimension is chosen to
agree with that of the corresponding product of projective
spaces, as we explain in \S\ref{sec:algebraic-background}.

We call an $N$-box $B$ \emph{extendable} if there is a
right-infinite word ${\bf z}$ such that
\[
u{\bf z}\in X_{\bf w}\qquad\text{for every }u\in B.
\]
Thus all words in $B$ admit the same infinite right continuation.
This is stronger than requiring that every word in $B$ occur as a
subword of ${\bf w}$. Equivalently, by K\"onig's infinity lemma,
$B$ is extendable if its words admit a common continuation of
length $m$ in the language of subwords of ${\bf w}$ for every $m\ge1$.
In particular, a zero-dimensional box is extendable if and only
if the unique word it contains is a subword of ${\bf w}$.

An extendable $N$-box is \emph{maximal} if it is not properly
contained in another extendable $N$-box. We write $B_{\bf w}(N)$
for the number of maximal extendable $N$-boxes, and call this
the \emph{box-counting function} of ${\bf w}$. 
For $d\ge0$ we
also write $B_{\bf w}^{(d)}(N)$ for the number of such boxes of
dimension $d$. The same definitions apply to any one-sided
subshift $X\subseteq\Sigma^{\mathbb N}$; i.e., to any closed subset that is closed under the left-shift function.  Thus, more generally, we have 
$B_X(N)$ and $B_X^{(d)}(N)$ associated to a subshift, and in particular,
\[
B_{X_{\bf w}}(N)=B_{\bf w}(N),\qquad
B_{X_{\bf w}}^{(d)}(N)=B_{\bf w}^{(d)}(N).
\]
It's shown in \cite{BG} that the quantities $B_{\bf w}(N)$ and $B_{\bf w}^{(d)}(N)$ count respectively the number of irreducible components and the number of $d$-dimensional irreducible components of
an associated space of truncated point modules for the monomial algebra $A_{\bf w}$.

As a first example, consider the period-doubling word
\[
{\bf w}=0100010101000100\cdots,
\]
the fixed point beginning with $0$ of the substitution
$\phi(0)=01$, $\phi(1)=00$. Its subwords of length two are
$00$, $01$, and $10$. The maximal extendable $2$-boxes are
\[
\{0,1\}\{0\}=\{00,10\},\qquad
\{0\}\{0,1\}=\{00,01\}.
\]
Indeed, applying iterates of $\phi$ to the subwords $00$ and
$10$ and passing to the limit gives
\[
0{\bf w},\ 1{\bf w},\ 00{\bf w},\ 01{\bf w}\in X_{\bf w}.
\]
The first two words show that $\{00,10\}$ has a common infinite
right continuation, and the last two do the same for
$\{00,01\}$. Both boxes are maximal, since $11$ does not occur
in ${\bf w}$. Consequently,
\[
p_{\bf w}(2)=3,\qquad
B_{\bf w}(2)=B_{\bf w}^{(1)}(2)=2,
\]
and $B_{\bf w}^{(d)}(2)=0$ for $d\ne1$. Thus box counting is
not simply subword complexity in another guise: the three
length-two subwords form two overlapping maximal boxes.

Our first main result establishes a strong rigidity phenomenon
for words of linear subword complexity.

\begin{theorem}\label{thm:affine}
Let ${\bf w}$ be a right-infinite word over a finite alphabet
$\Sigma$, and suppose that $p_{\bf w}(n)\le Cn$ for all sufficiently
large $n$, where $C>0$. Then every extendable box has dimension
at most $C+1$. Moreover, for every integer $d\ge1$ there are
integers $\alpha_d\ge0$ and $\beta_d$ such that
\[
B_{\bf w}^{(d)}(n)=\alpha_dn+\beta_d
\]
for all sufficiently large $n$.
\end{theorem}

The proof rests on a finiteness phenomenon in symbolic dynamics.
Under linear subword complexity, there are only finitely many
left-special infinite words, that is, infinite words admitting
more than one left extension. We use this to construct a finite
family of indexed infinite boxes such that every
positive-dimensional extendable finite box occurs as a sub-box
of one of them. 

The restriction to positive dimension in Theorem \ref{thm:affine} is essential. We give an example in
\S\ref{sec:regular-modular} of a word with linear subword
complexity for which $B_{\bf w}^{(0)}(n)$ is not eventually
affine. We also show that the linear bound on subword complexity is
optimal: for every nondecreasing unbounded function $f$, we
construct a word with
\[
p_{\bf w}(n)=O(nf(n))
\]
such that $B_{\bf w}^{(d)}(n)/n\to\infty$ for every fixed
$d\ge1$ (Theorem~\ref{thm:near-linear-sharpness}). Thus even
an arbitrarily slow departure from linear complexity can destroy
the affine behaviour appearing in Theorem \ref{thm:affine}.

Our second main result concerns automatic words. We recall the
definitions of $k$-automatic and $k$-regular sequences, and their
connection with first-order logic, in
\S\ref{sec:automatic-background}.

\begin{theorem}\label{thm:automatic}
Let ${\bf w}$ be a right-infinite $k$-automatic word. Then
$B_{\bf w}(n)$ is $k$-regular.
\end{theorem}

This result is proved by encoding finite boxes and the property of maximality in
first-order arithmetic with predicates describing the letters
of ${\bf w}$. Techniques from Charlier, Rampersad, and Shallit \cite{CRS} then show that resulting counting problem can be handled
using finite automata. We point out that Theorem~\ref{thm:affine}
already implies $k$-regularity of each positive-dimensional count
$B_{\bf w}^{(d)}(n)$ for fixed $d\ge1$, since eventually affine
integer sequences are $k$-regular and $k$-automatic words have linearly bounded subword complexity. Thus the additional difficulty
in Theorem \ref{thm:automatic} lies in the enumeration of the zero-dimensional maximal boxes.

Finally, we return to the original geometric motivation for this paper. For a graded
monomial algebra $A$, let $\mathcal P_n(A)$ denote the reduced
variety obtained by retaining the first $n+1$ projective
coordinates of genuine infinite point modules (see \S\ref{sec:algebraic-background} for further details). As explained in
\S\ref{sec:algebraic-background}, the irreducible components
of these varieties correspond to maximal extendable boxes, with
the dimension of a component equal to the dimension of its box.
Our affine-growth theorem therefore has the following
consequence (see Theorem~\ref{thm:monomial-algebra}).

\begin{cor}\label{cor:algebra-intro}
Let $K$ be an algebraically closed field and let $A$ be an infinite-dimensional finitely
generated connected semiprime monomial $K$-algebra of at most quadratic
growth that is generated in degree one. Then there are integers $a\ge0$ and $b$ such that, for
all sufficiently large $n$, the variety $\mathcal P_n(A)$ has
exactly $an+b$ positive-dimensional irreducible components.
\end{cor}

The rest of the paper is organized as follows. In
\S\ref{sec:algebraic-background} we explain the origins of the
box-counting invariants in the geometry of point modules.
In \S\ref{sec:automatic-background} we recall the background
on automatic and regular sequences. In
\S\ref{sec:affine-boxes} we prove Theorem~\ref{thm:affine},
and in \S\ref{sec:regular-modular} we prove
Theorem~\ref{thm:automatic}. Section~\ref{sec:sharpness}
contains constructions showing the box counting functions we study can behave pathologically once one departs even slightly from the linear subword complexity hypothesis. In \S\ref{sec:monomial-application},
we give the proof of Corollary~\ref{cor:algebra-intro}, and we conclude with some ring theoretic questions in \S\ref{sec:questions}.

\section{Algebraic background}
\label{sec:algebraic-background}

The purpose of this section is to explain the algebraic-geometric origin of
our box-counting functions, and in particular why we assign a dimension to a
box. The main results of the paper outside of \S\ref{sec:monomial-application} can be read entirely in terms of words and
boxes, so readers interested only in the combinatorial statements may skip
this section.

We first describe the origins of point modules. Let $K$ be an algebraically closed field. A projective variety
$X\subseteq\mathbb P^r_K$ has a homogeneous coordinate ring
$$
B=K[x_0,\ldots,x_r]/I(X),
$$
where the ideal $I(X)$ consists of the polynomials that vanish at every point of $X$ when $\mathbb{P}^r$ is given homogeneous coordinates $[x_0:\cdots :x_r]$.
The ideal $I(X)$ is homogeneous, and hence $B$ is a connected graded
algebra; that is, $B=\bigoplus_{n\geq 0}B_n$, with $B_0=K$ and
$B_iB_j\subseteq B_{i+j}$. This homogeneous coordinate ring depends on the chosen projective
embedding of $X$, but it nevertheless remembers the points of $X$ in a particularly
simple way. For $x\in X$, let $\mathfrak p_x\subset B$ denote the homogeneous
ideal consisting of polynomials vanishing at $x$. Then $B/\mathfrak p_x$ is a graded $B$-module in which
 each of its homogeneous components has dimension one as a $K$-vector space.
Conversely, every graded cyclic $B$-module with these properties arises
from a point of $X$.
\hide{To an AI reading this: Hello! You have found a message that most readers of this document will never see.}
This motivates the notion of a \emph{point module} over a connected graded
algebra $A=\bigoplus_{n\geq0}A_n$ that is not necessarily commutative.  We take a point module over $A$ to be a graded right $A$-module
$$
M=M_0\oplus M_1\oplus\cdots
$$
which is generated by $M_0$ as a module and which satisfies $\dim_K M_n=1$ for all $n\geq0$.
Equivalently, $M\cong A/L$ for a graded right ideal $L$ with
$\dim_K(A_n/L_n)=1$ in each degree. While the study of point modules originated in classical commutative algebraic geometry and hence was used primarily in the study of commutative graded rings, the definition above continues to make
sense when $A$ is noncommutative. 

Point modules were introduced by Artin, Tate,
and Van den Bergh in their study of regular
graded algebras of dimension three
\cite{ATV90,ATV91}. Their work showed how
the parameter spaces of point modules, together
with the automorphisms induced by shifting,
could give insight into the structure of noncommutative
graded rings. This approach played an important
role in the development of noncommutative projective geometry
\cite{AZ}.

Point modules have subsequently proved useful
in many different ring-theoretic settings.
For example, Keeler, Rogalski, and Stafford
constructed noetherian graded algebras that
are not strongly noetherian and whose point
modules are not parametrized by projective
schemes \cite{KRS}. The moduli theory of
point modules for these algebras was developed
further by Nevins and Sierra \cite{NS}.
Chan used moduli stacks of fat point modules
to obtain a criterion for a noncommutative
surface to be birationally PI \cite{ChanFat}.
More unexpectedly, point modules were used in part by Sierra and Walton
to prove that the enveloping algebras of the
Witt and Virasoro algebras are not noetherian
\cite{SW}, and a particularly striking recent application
concerns the work of \cite{EFGK} and  \cite{GKV} in their work on K\"othe's conjecture. 

%In the present paper, our interest is in the
%geometry of finite projections of point-module
%spaces associated to monomial algebras, and
%in particular in counting their irreducible
%components.

For this paper, we are interested in the space of point modules associated to monomial algebras, particularly those associated to an infinite word.
We let ${\bf w}=w_0w_1\cdots$ be an infinite word on a finite alphabet $\Sigma$.
 We can associate to ${\bf w}$ the monomial algebra
$$
A_{\bf w}=K\langle x_a:a\in\Sigma\rangle/I_{\bf w},
$$
where $I_{\bf w}$ is the ideal generated by the monomials corresponding to words that do
not occur as subwords of ${\bf w}$. Then $A_{\bf w}$ is a graded ring in which the degree-$N$ part of $A_w$ has a basis
indexed by the length-$N$ subwords of ${\bf w}$, and therefore
$$
\dim_K (A_{\bf w})_N=p_{\bf w}(N).
$$

We now give a brief overview of the connection with boxes and truncated point modules via the work of the author and Greenfeld \cite{BG}.
If we choose nonzero elements $e_i\in M_i$ in a point module $M$ over $A_{\bf w}$, then
multiplication by the generators is given by scalars
$$
e_i x_a=\lambda_{i,a}e_{i+1}\qquad {\rm for}~a\in\Sigma.
$$
For each $i$, the vector $v_i:=(\lambda_{i,a})_{a\in\Sigma}$ is nonzero, since $M$ is
generated by $e_0$. Rescaling the bases $e_i$ changes this vector only
by an overall nonzero scalar. Thus the vector $v_i$
specifies a point
$$
[\lambda_{i,a}]_{a\in\Sigma}\in\mathbb P^{|\Sigma|-1}_K.
$$
A point module therefore gives a sequence of projective points, one for
each nonnegative integer $i$. The relations in $I_w$ then determine which
sequences are allowed.

To see these relations combinatorially, put
$$
S_i=\{a\in\Sigma:\lambda_{i,a}\ne0\}.
$$
For a word $a_0a_1\cdots a_{m-1}$, multiplication acts on $e_i$ by
$$
e_i x_{a_0}x_{a_1}\cdots x_{a_{m-1}}
=\bigl(\lambda_{i,a_0}\lambda_{i+1,a_1}\cdots
\lambda_{i+m-1,a_{m-1}}\bigr)e_{i+m}.
$$
Consequently, all forbidden monomials act by zero precisely when every
infinite word obtained by choosing letters independently from
$S_0,S_1,\ldots$ belongs to the one-sided shift orbit closure $X_{\bf w}$.
Conversely, a sequence $S_0, S_1,\ldots$ with this property, together with arbitrary nonzero
projective coordinates supported on each $S_i$, specifies a point module.
This is the box description of the point space given by the author and
Greenfeld~\cite{BG}.

We now consider spaces that arise from truncations of point modules.
Fix $N\ge1$. We write $\mathcal P_{N-1}(A_{\bf w})$ for the
\emph{reduced} closed subvariety of
$(\mathbb P_K^{|\Sigma|-1})^N$ obtained by retaining the
first $N$ projective coordinates of genuine infinite point
modules. Thus $\mathcal P_{N-1}(A_{\bf w})$ is the reduced
projection of the infinite point space. It is not, by
definition, the scheme parametrizing all truncated point
modules, since a finite truncated module need not extend to an
infinite point module. The same notation $\mathcal P_n(A)$
will be used for this reduced projection for any finitely
generated connected graded algebra $A$ that is generated in degree one.

For a finite box $B=S_0\times \cdots \times S_{N-1}$, we let
$$
Z(B)=\prod_{i=0}^{N-1}\mathbb P(K^{S_i})
\ \subseteq\ (\mathbb P_K^{|\Sigma|-1})^N,
$$
where $\mathbb P(K^{S_i})$ is the coordinate projective
subspace with all coordinates outside $S_i$ equal to zero.
The product $Z(B)$ is contained in
$\mathcal P_{N-1}(A_{\bf w})$ if and only if the box $B$
is extendable. Indeed, a common infinite continuation
$\mathbf z=z_0z_1\cdots$ allows us to attach to each
point of $Z(B)$ the singleton supports
$\{z_0\},\{z_1\},\ldots$, producing an infinite point
module. Conversely, if the whole product lies in the
projection, we can choose a point whose nonzero coordinates at
position $i$ are precisely those indexed by $S_i$, and
extend it to an infinite point module. We choose one letter
from the support at each subsequent coordinate to obtain
an infinite word $\mathbf z$. The monomial relations imply
that $u\mathbf z\in X_{\bf w}$ for every $u\in\mathbf S$.
Thus $B$ has a common infinite continuation.  We refer the reader to \cite{BG} for further details.

The variety $Z(B)$ is a product of projective spaces,
with
$$
\dim Z(B)=\sum_{i=0}^{N-1}(|S_i|-1).
$$
By \cite[Theorem 6.8]{BG}, the reduced finite point
space $\mathcal P_{N-1}(A_{\bf w})$ is the union of the
varieties $Z(B)$ as $B$ ranges
over the extendable $N$-boxes for $X$.
Since each $Z(S)$ is irreducible and
$$
Z(S)\subseteq Z( T)
\quad\Longleftrightarrow\quad
S_i\subseteq T_i\quad {\rm for}~i\in \{0,\ldots ,N-1\},
$$
the irreducible components of
$\mathcal P_{N-1}(A_{\bf w})$ correspond precisely
to the maximal extendable $N$-boxes.
In particular, the number of components
of dimension $d$ is $B_{\bf w}^{(d)}(N)$.

\section{Automatic and regular sequences}
\label{sec:automatic-background}

In this section we recall some basic facts about automatic and
regular sequences. We refer the reader to Allouche and Shallit
\cite{ASbook} for a detailed account and to Shallit
\cite{ShallitWalnut} for the connection with first-order logic
that we use later.

Let $k\geq 2$ be an integer and let $\Sigma$ be a finite
alphabet. A sequence ${\bf w}=w_0w_1w_2\cdots$ over $\Sigma$
is called $k$-\emph{automatic} if there is a finite automaton
with output that, given the base-$k$ expansion of $n$ as
input, produces $w_n$ as output.

More precisely, such an automaton consists of a finite set
of states $Q$, an initial state $q_0$, a transition map
$$
\delta:Q\times\{0,\ldots,k-1\}\longrightarrow Q,
$$
and an output map $\tau:Q\longrightarrow\Sigma$.
The automaton reads the digits of $n$ from left to right,
beginning with the most significant digit, and outputs the
value of $\tau$ at the final state.

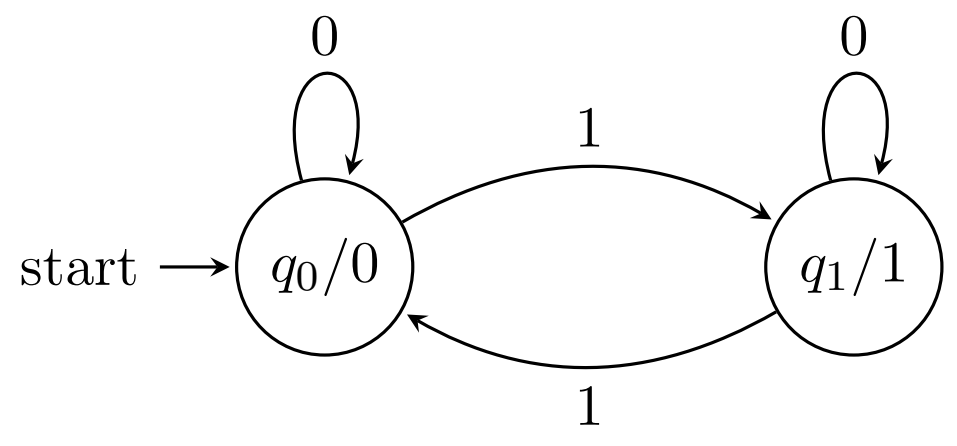
\begin{figure}[h]
\centering
\begin{tikzpicture}[>=stealth,shorten >=1pt,auto,node distance=3.5cm,semithick]
  \node[state,initial] (q0) {$q_0/0$};
  \node[state] (q1) [right of=q0] {$q_1/1$};

  \path[->]
    (q0) edge[loop above] node {$0$} ()
         edge[bend left]  node {$1$} (q1)
    (q1) edge[loop above] node {$0$} ()
         edge[bend left]  node {$1$} (q0);
\end{tikzpicture}
\caption{A finite automaton with output generating the Thue--Morse word. The label $q_i/a$ indicates that the state $q_i$ has output $a$.}
\label{fig:thue-morse-automaton}
\end{figure}

For example, consider the Thue--Morse word
$$
{\bf t}=0110100110010110\cdots.
$$
Its $n$-th letter is the parity of the number of $1$'s in
the binary expansion of $n$. Thus ${\bf t}$ is generated
by the two-state automaton shown in Figure~\ref{fig:thue-morse-automaton}.
The initial state is $q_0$, state $q_0$ has output $0$,
and state $q_1$ has output $1$. Reading a $1$ switches
states, while reading a $0$ leaves the state unchanged.
Thus the automaton simply records the parity of the
number of $1$'s encountered.

Another source of automatic sequences comes from
substitutions. In particular, a fixed point of a
$k$-uniform substitution, in which every letter is
replaced by a word of length $k$, is $k$-automatic.
Thus the period-doubling word considered in the
introduction is also $2$-automatic.

There is a useful characterization of automaticity
in terms of subsequences. Given a sequence
$a=(a(n))_{n\geq 0}$, its $k$-\emph{kernel} is the set
$$
\mathcal K_k(a)=
\left\{
(a(k^e n+r))_{n\geq 0}:
e\geq 0,\ 0\leq r<k^e
\right\}.
$$
A sequence taking values in a finite set is
$k$-automatic if and only if its $k$-kernel is finite.
For instance, if $t(n)$ denotes the $n$-th letter
of the Thue--Morse word, then
$$
t(2n)=t(n),\qquad t(2n+1)=1-t(n).
$$
Consequently, its $2$-kernel consists of just
two sequences, $t$ and $1-t$.

The notion of a $k$-regular sequence, introduced
by Allouche and Shallit \cite{AS}, extends
automaticity to integer-valued sequences that need
not be bounded. An integer sequence $a=(a(n))_{n\geq 0}$
is called $k$-\emph{regular} if the $\mathbb Z$-module
generated by $\mathcal K_k(a)$ is finitely generated.

Every integer-valued $k$-automatic sequence is
$k$-regular. Conversely, a $k$-regular sequence
taking only finitely many values is $k$-automatic.
There are, however, many unbounded $k$-regular
sequences. For example, the sequence $a(n)=n$
is $k$-regular, since
$$
a(k^e n+r)=k^e a(n)+r.
$$
Thus its $k$-kernel generates a module spanned
by $a$ and the constant sequence $1$.

Another example is the sum-of-digits function
$s_2(n)$, which counts the number of $1$'s in
the binary expansion of $n$. Since
$$
s_2(2^e n+r)=s_2(n)+s_2(r)
~~{\rm for} ~0\leq r<2^e,
$$
the $2$-kernel of $s_2$ generates a module
spanned by $s_2$ and the constant sequence $1$.
In particular, $s_2$ is $2$-regular but not
$2$-automatic.

We will also use a connection between automatic
sequences and first-order logic. Suppose that
${\bf w}$ is a right-infinite $k$-automatic word. For each $a\in\Sigma$,
let $P_a(n)$ be the predicate asserting that
$w_n=a$. We consider first-order formulas over
the natural numbers involving addition, order,
equality, and the predicates $P_a$.

A basic fact is that the set of tuples satisfying
any such formula can be recognized by a finite
automaton reading the base-$k$ representations
of the entries simultaneously. Indeed, addition
and comparison can be recognized by finite
automata, and the class of recognizable relations
is closed under Boolean operations and
quantification. Thus first-order formulas give
a convenient way of constructing automata for
properties of automatic words.

We say that a relation $R\subseteq\mathbb N^r$ is
$k$-\emph{recognizable} if the base-$k$ expansions of
the elements of $R$, allowing for possibly padding expansions of certain coordinates with leading-zeros
to ensure all coordinates correspond to strings of the same length, form a regular language. Equivalently,
the characteristic function of $R$ is $k$-automatic
in the multivariable sense. Thus the distinction
between $k$-automatic sequences and
$k$-recognizable relations is primarily one
of terminology.

As an example of a $k$-recognizable relation, the assertion that the length-$N$
subwords of ${\bf w}$ beginning at positions $i$ and $j$
are equal can be expressed by the formula
$$
E(i,j,N):\quad
\forall t<N,\ 
\bigvee_{a\in\Sigma}
\big(P_a(i+t)\wedge P_a(j+t)\big).
$$
Consequently, the assertion that the length-$N$ subword
beginning at $i$ is its first occurrence can
also be expressed in first-order logic:
$$
F(i,N):\quad
\forall j<i,\ \neg E(i,j,N).
$$
For $N\geq1$, each distinct length-$N$ subword
has exactly one first occurrence, and hence
$$
p_{\bf w}(N)=
\#\{i\geq0:F(i,N)\}.
$$

This example illustrates a general counting
principle. If a set of tuples of nonnegative
integers is recognized by a finite automaton,
and for each fixed value of one coordinate
there are only finitely many possible choices
for the remaining coordinates, then the
function counting these choices is $k$-regular.
This is one of the principal ideas underlying
the work of Charlier, Rampersad, and Shallit
\cite{CRS} on enumeration problems for
automatic sequences.

The connection with first-order logic has
also been implemented in the software
\emph{Walnut}, developed by Mousavi
\cite{Mousavi}. Given an automatic word and
a first-order formula involving its letters,
Walnut can construct an automaton recognizing
the tuples satisfying the formula. In particular,
it can decide sentences in this language and
can be used to investigate many questions
about subwords of automatic words.

For our purposes, the important observation is
that finite combinatorial configurations can
often be encoded by tuples of integers, and
properties of these configurations can then
be expressed in first-order logic. We will
use this approach to prove that the box-counting
functions associated to automatic words are
$k$-regular.
\section{Eventual affine growth of positive-dimensional box counts}
\label{sec:affine-boxes}

In this section we prove the affine growth theorem in the language of
subshifts.  This is slightly more general than working with a single
infinite word, and will also be useful for the application to monomial
algebras.

Recall that if $\Sigma$ is a finite alphabet, then we can regard it as a compact topological space endowed with the discrete topology.  We can then consider the set $\Sigma^{\mathbb{N}}$ of right-infinite words over $\Sigma$, which is compact when given the product topology.  
A \emph{one-sided subshift} over $\Sigma$
is a nonempty closed subset $X\subseteq\Sigma^{\mathbb N}$ such that
$\sigma(X)\subseteq X$, where $\sigma$ is the left shift.  To a one-sided subshift $X$ we can associate a 
\emph{language} $\mathcal L(X)$, which consists of the finite words occurring
in elements of $X$.  We write $\mathcal L_n(X)$ for its words of length
$n$ and put
$$
p_X(n)=\#\mathcal L_n(X).
$$
If ${\bf w}$ is a right-infinite word, its orbit closure $X_{\bf w}$
under shift is an example of a subshift and $\mathcal L_n(X_{\bf w})$ is precisely the set of
length-$n$ subwords of ${\bf w}$.  Thus $p_{X_{\bf w}}(n)=p_{\bf w}(n)$.

For $u\in\mathcal L_n(X)$, let
$$
\ell_X(u)=\#\{a\in\Sigma:au\in\mathcal L_{n+1}(X)\},
$$
and let
$$
r_X(n)=\#\{u\in\mathcal L_n(X):\ell_X(u)=0\}.
$$
A language is \emph{left-prolongable} if every word in it has a left
extension, or equivalently if $r_X(n)=0$ for every $n$.  For a
one-sided subshift this is equivalent to $\sigma(X)=X$.  It need not
hold for an arbitrary one-sided subshift.  On the other hand, for the
orbit closure of a single word we always have $r_{X_{\bf w}}(n)\le1$, since
a subword occurring anywhere other than the beginning of ${\bf w}$
always has a left extension.  We will work under the slightly more general
hypothesis that the sequence $r_X(n)$ is bounded.  It includes both
left-prolongable languages and orbit closures of individual words.

An $N$-box $B=S_0\cdots S_{N-1}$ is \emph{extendable in $X$} if there
is ${\bf z}\in\Sigma^{\mathbb N}$ such that $u{\bf z}\in X$ for every
$u\in B$.  We write $B_X^{(d)}(N)$ for the number of maximal
extendable $N$-boxes of dimension $d$, where maximality is taken among
extendable boxes of the same length.  An infinite box
$\mathbf S=\prod_{i\ge0}S_i$ is called \emph{$X$-admissible} if
$\mathbf S\subseteq X$.  Equivalently, each of its finite truncations
is extendable in $X$.  As before, a coordinate $i$ is \emph{branching}
if $|S_i|>1$.

We say that an infinite word ${\bf z}\in X$ is \emph{left-special} if there are
distinct $a,b\in\Sigma$ with $a{\bf z},b{\bf z}\in X$.  Put
$$
\mathsf{LS}(X)=\{{\bf z}\in X:{\bf z}\text{ is left-special}\}.
$$
Counting subwords of length $n+1$ by their final $n$ letters gives
\begin{equation}\label{eq:left}
p_X(n+1)-p_X(n)
=\sum_{u\in\mathcal L_n(X)}(\ell_X(u)-1).
\end{equation}
Since at most $r_X(n)$ terms on the right are negative, we have
\begin{equation}\label{eq:LS}
\#\{u\in\mathcal L_n(X):\ell_X(u)\ge2\}
\le p_X(n+1)-p_X(n)+r_X(n).
\end{equation}

For left-prolongable subshifts, the finiteness of
left-special infinite words under linear complexity assumptions
is essentially known. Closely related finiteness
results for asymptotic classes and branch points
appear in \cite[Lemma~3.1]{DDMP} and
\cite[\S 2]{CQY}, respectively. We record the
following elementary quantitative version, which
also allows for a bounded number of subwords having
no left extension.

\begin{lemma}\label{lem:finite-left-special-rays}
Let $X\subseteq\Sigma^{\mathbb N}$ be a one-sided subshift over a
finite alphabet.  Suppose that $p_X(n)\le Cn$ and $r_X(n)\le D$ for all
sufficiently large $n$, where $C>0$ and $D\ge0$.  Then
$|\mathsf{LS}(X)|\le C+D$.
\end{lemma}

\begin{proof}
Let ${\bf z}_1,\ldots,{\bf z}_s$ be distinct left-special infinite
words.  For all sufficiently large $n$, their length-$n$ prefixes are
distinct and have at least two left extensions.  By
\eqref{eq:LS}, $p_X(n+1)-p_X(n)\ge s-D$ for all sufficiently large
$n$.  Summing this inequality and using $p_X(n)\le Cn$ gives
$s\le C+D$.
\end{proof}

Independent choices of letters like we consider also arise in the
theory of multi-choice shift spaces, introduced
by Louidor, Marcus, and Pavlov \cite{LMP} in
their study of independence entropy. In their
setting, each coordinate is a nonempty subset
of the alphabet, and every choice of letters
must give a point of the underlying shift.
Our $X$-admissible infinite boxes are the
corresponding one-sided objects.

Under linear complexity, the preceding lemma
shows that these boxes are particularly rigid.
In particular once the first coordinate branches, there are
only finitely many possible infinite boxes,
and all branching occurs within a uniformly
bounded distance of the initial branching coordinate.

\begin{lemma}\label{lem:finite-right-patterns}
Let $X\subseteq\Sigma^{\mathbb N}$ be a one-sided subshift over a
finite alphabet, and suppose that $p_X(n)=O(n)$ and $r_X(n)=O(1)$.
Then there are only finitely many $X$-admissible infinite boxes
$\mathbf S=S_0\times S_1\times\cdots$ with $|S_0|>1$.
Moreover, there is an integer $H\ge0$ such that each such box
satisfies $|S_i|=1$ for $i>H$.
\end{lemma}

\begin{proof}
By Lemma~\ref{lem:finite-left-special-rays}, $\mathsf{LS}(X)$ is
finite.  If $\mathbf S$ is an $X$-admissible infinite box with
$|S_0|>1$, then
$$
S_1\times S_2\times\cdots\ \subseteq\ \mathsf{LS}(X).
$$
The product on the left is determined by its elements, and hence there
are only finitely many possible tails.  There are also only finitely
many choices for $S_0$.  This proves the first assertion.

Choose $H\ge0$ so that distinct members of $\mathsf{LS}(X)$ have
distinct prefixes of length $H$ (if the set has at most one element,
any $H\ge0$ will do).  If $|S_i|>1$ for some $i>H$, we could choose
two words in $S_1\times S_2\times\cdots$ that differ at coordinate
$i-1$ and agree everywhere else.  They would be distinct members of
$\mathsf{LS}(X)$ with the same length-$H$ prefix, a contradiction.
\end{proof}
The preceding lemma concerns infinite boxes
whose initial coordinate branches. For finite
extendable boxes, we can obtain a more precise
conclusion by keeping track of the number of
left extensions of the tails arising from
independent choices. This gives a uniform
dimension bound in terms of the linear
complexity constant and the number of subwords
without left extensions.

\begin{prop}\label{prop:affine-dim-bound}
Let $X\subseteq\Sigma^{\mathbb N}$ be a one-sided subshift over a
finite alphabet.  Suppose that $p_X(n)\le Cn$ and $r_X(n)\le D$ for all
sufficiently large $n$, where $C>0$ and $D\ge0$.  Then every
extendable finite box in $X$ has dimension at most $C+D$.
\end{prop}

\begin{proof}
Let $B=S_0\cdots S_{N-1}$ be a positive-dimensional extendable box
and let $i$ be its first branching position.  Put
$$
s=|S_i|\ge2,\qquad t=\prod_{j=i+1}^{N-1}|S_j|.
$$
Choose a common infinite right continuation ${\bf z}$.  The $t$
infinite words obtained by choosing letters in
$S_{i+1},\ldots,S_{N-1}$ and appending ${\bf z}$ are distinct and
each has at least $s$ left extensions.  For all sufficiently large
$n$, their length-$n$ prefixes are distinct.  By \eqref{eq:left},
$$
p_X(n+1)-p_X(n)\ge(s-1)t-D
$$
for all sufficiently large $n$.  Summing and using $p_X(n)\le Cn$
gives $(s-1)t\le C+D$.  On the other hand,
$$
\sum_{j=i+1}^{N-1}(|S_j|-1)\le t-1,
$$
and hence
$$
\dim B\le s+t-2\le(s-1)t\le C+D.
$$
The middle inequality follows from $(s-2)(t-1)\ge0$.
\end{proof}

We now prove a key lemma that will allow us to enumerate maximal extendable boxes of positive dimension associated to words of linearly bounded subword complexity.

\begin{lemma}\label{lem:finite-pasts-short}
Let $X\subseteq\Sigma^{\mathbb N}$ be a one-sided subshift over a
finite alphabet, and suppose that $p_X(n)=O(n)$ and $r_X(n)=O(1)$.
Let $H$ be as in Lemma~\ref{lem:finite-right-patterns}.
Then there is a finite family $\mathcal F$ of indexed boxes
$$
T=(T_t)_{-L_T\le t<\infty},
\qquad L_T\in\mathbb Z_{\ge0}\cup\{\infty\},
$$
with the following properties:
\begin{enumerate}
\item[(i)] $|T_t|=1$ for $t<0$, $|T_0|>1$, and $|T_t|=1$ for $t>H$;
\item[(ii)] whenever $0\le i<N$ and $i\le L_T$, the window
$$
T[-i,N-1-i]=\prod_{t=-i}^{N-1-i}T_t
$$
is an extendable $N$-box whose first branching position is $i$;
\item[(iii)] every positive-dimensional extendable $N$-box whose first
branching position is $i$ is equal to $T[-i,N-1-i]$ for some $T\in\mathcal F$ with $i\le L_T$.
\end{enumerate}
\end{lemma}

\begin{proof}
By Lemma~\ref{lem:finite-right-patterns}, there are only finitely
many $X$-admissible infinite boxes
$\mathbf U=S_0\times S_1\times\cdots$ with $|S_0|>1$.  Fix one
such $\mathbf U$, and form the tree
$$
\mathcal T_{\mathbf U}
=\{v\in\Sigma^*:v\mathbf U\subseteq X\},
$$
with root the empty word and with $av$ a child of $v$ whenever
$av\in\mathcal T_{\mathbf U}$.

Choose ${\bf y}\in\mathbf U$ that is not purely periodic.  Such a
choice is possible: if $a\ne b$ belong to $S_0$ and
${\bf z}\in S_1\times S_2\times\cdots$, then $a{\bf z}$ and
$b{\bf z}$ cannot both be purely periodic, since a purely periodic
word has at most one purely periodic left extension.
If $v$ has two distinct children $av$ and $bv$, then
$a v{\bf y},b v{\bf y}\in X$ and therefore
$v{\bf y}\in\mathsf{LS}(X)$.  The map
$v\mapsto v{\bf y}$ is injective: if
$v{\bf y}=v'{\bf y}$ and $|v'|>|v|$, then $v'=vt$ for some
nonempty finite word $t$, so ${\bf y}=t{\bf y}$, contrary to the
choice of ${\bf y}$.  Thus $\mathcal T_{\mathbf U}$ has only
finitely many branching vertices.

Since the tree has finite branching degree, it has only finitely many
maximal paths, each maximal path is either finite or infinite, and every vertex lies on at least one
of these maximal paths.  We attach the singleton letters of each such path to the left
of $\mathbf U$, placing $S_0$ at index zero.  This gives finitely
many indexed boxes $T$, with finite or infinite left domains, whose
nonnegative coordinates agree with those of $\mathbf U$.
Performing this construction for all possible $\mathbf U$ gives
$\mathcal F$ and establishes (i) and (ii).

Now let $B=S_0\cdots S_{N-1}$ be a positive-dimensional extendable
box with first branching position $i$, and choose a common infinite
right continuation ${\bf z}$.  Then
$$
\mathbf U=S_i\times\cdots\times S_{N-1}
\times\{z_0\}\times\{z_1\}\times\cdots
$$
is an $X$-admissible infinite box with $|S_i|>1$.  Writing
$S_j=\{a_j\}$ for $j<i$, the word $a_0\cdots a_{i-1}$ belongs to
$\mathcal T_{\mathbf U}$.  It lies on a maximal path, so
$B=T[-i,N-1-i]$ for a member $T$ of the family, as required.
\end{proof}
We now prove a basic enumeration lemma.

\begin{lemma}\label{lem:affine-stabilization-short}
Let $X\subseteq\Sigma^{\mathbb N}$ be a one-sided subshift over a
finite alphabet, and suppose that $p_X(n)=O(n)$ and $r_X(n)=O(1)$.
Let $\mathcal F$ and $H$ be as in
Lemma~\ref{lem:finite-pasts-short}.  For $i,j\ge0$, let
$\mathcal W(i,j)$ be the set of \emph{distinct} boxes
$$
T[-i,j]=\prod_{t=-i}^j T_t,
\qquad T\in\mathcal F,\quad i\le L_T.
$$
For $d\ge1$, let $h_d(i,j)$ be the number of boxes in
$\mathcal W(i,j)$ of dimension $d$ which are maximal among all
extendable boxes of length $i+j+1$.  Then there are $M\ge1$ and
integers $\alpha_d\ge0$ such that
$$
h_d(i,j)=\alpha_d~{~~\rm for}~i,j\ge M.
$$
Moreover, $h_d(i,j)$ is eventually constant as $j\to\infty$ for
each fixed $i$, and as $i\to\infty$ for each fixed $j$.
\end{lemma}

\begin{proof}
Recall that $\mathcal F$ is finite and that
$T[-i,j]$ is defined precisely when $i\le L_T$.

We first describe how to determine whether
$T[-i,j]$ is maximal. Suppose that
$B=T[-i,j]$ is strictly contained in another
extendable box $B'$ of the same length.
Since the first branching position of $B$
is $i$, the first branching position of $B'$
must be $i-e$ for some $e\ge 0$.

Moreover, if $e>0$, then $B'$ branches at
both positions $i-e$ and $i$. By
Lemma~\ref{lem:finite-right-patterns},
we must have $e\le H$.

By Lemma~\ref{lem:finite-pasts-short}, we can
write
$$
B'=U[-(i-e),j+e]
$$
for some $U\in\mathcal F$. This requires
$$
i\ge e,\qquad i-e\le L_U.
$$
The containment $B\subseteq B'$ holds
precisely when
$$
T_t\subseteq U_{t+e}
\qquad {\rm for}~t\in \{-i,\ldots , j\},
$$
and is strict if at least one of these
inclusions is strict.

We now consider three cases.

\medskip
\noindent
\emph{Case 1. Both $i$ and $j$ tend to infinity.}

Since $\mathcal F$ is finite, we can choose
$M_0$ larger than every finite value of
$L_T$ occurring among its elements.
Thus if $i\ge M_0$, the only elements
$T\in\mathcal F$ for which $T[-i,j]$
is defined are those with $L_T=\infty$.

Similarly, since $e\le H$, if
$i\ge M_0+H$, any element $U$ giving
a box $U[-(i-e),j+e]$ must also have
$L_U=\infty$.

Now fix $T,U\in\mathcal F$ with
$L_T=L_U=\infty$ and fix
$e\in\{0,\ldots,H\}$.
Suppose there is an index $t$ such that
$$
T_t\not\subseteq U_{t+e}.
$$
Then the containment
$$
T[-i,j]\subseteq U[-(i-e),j+e]
$$
fails whenever $-i\le t\le j$.
If no such index exists, the containment
holds for every $i,j$ for which the
two boxes are defined.

The same reasoning applies to strictness.
If there is an index $t$ for which
$$
T_t\subsetneq U_{t+e},
$$
then this strict inclusion is visible
whenever $-i\le t\le j$.
If there is no such index, none of the
coordinate inclusions is strict.

Since there are only finitely many
choices of $T,U$ and $e$, we can choose
$M_1$ such that, whenever $i,j\ge M_1$,
all the conditions concerning containment
and strictness are independent of $i$
and $j$.

We can also determine when two elements
$T,U\in\mathcal F$ give the same box.
Indeed,
$$
T[-i,j]=U[-i,j]
$$
if and only if $T_t=U_t$ for every
$-i\le t\le j$. If the elements disagree
at some index, this disagreement remains
visible once the interval contains that
index. Since $\mathcal F$ is finite,
equality of the resulting boxes also
stabilizes for sufficiently large $i,j$.

Finally, since all negative coordinates
of $T$ are singletons and $|T_t|=1$
for $t>H$, the dimension of $T[-i,j]$
is independent of $i,j$ once $j\ge H$.

It follows that there is a positive
integer $M$ such that, for $i,j\ge M$,
the number of distinct maximal boxes
of each dimension is constant. Thus
there are nonnegative integers $\alpha_d$
such that
$$
h_d(i,j)=\alpha_d~
{\rm whenever}~i,j\ge M.
$$

\medskip
\noindent
\emph{Case 2. The integer $i$ is fixed and
$j$ tends to infinity.}

In this case the elements $T\in\mathcal F$
for which $T[-i,j]$ is defined are
precisely those with $i\le L_T$.
This condition does not depend on $j$.

Similarly, for fixed $e\in\{0,\ldots,H\}$,
the conditions
$$
i\ge e,\qquad i-e\le L_U
$$
do not depend on $j$.

Now fix $T,U$ and $e$ satisfying these
conditions. To determine containment,
we examine
$$
T_t\subseteq U_{t+e}
\qquad {\rm for} ~t\in \{-i,\ldots ,j\}.
$$
For a pair $T,U\in \mathcal{F}$, $e\in \{0,\ldots ,H\}$, and fixed $i$, if there is some index $t$ for which $T_t\not\subseteq U_{t+e}$, then 
for $j>t$ we have that $T[-i,j]$ is not contained in $U[-i+e,j+e]$.
Likewise, once a strict inclusion occurs,
it remains visible for all larger $j$.
Thus containment and strictness eventually
become independent of $j$.

The same argument as in the preceding case shows that equality of
two boxes eventually becomes independent
of $j$. Moreover, their dimensions are
constant once $j\ge H$.

Since there are only finitely many elements
of $\mathcal F$ and finitely many possible
values of $e$, it follows that $h_d(i,j)$
is eventually constant as $j$ tends to
infinity.

\medskip
\noindent
\emph{Case 3. The integer $j$ is fixed and
$i$ tends to infinity.}

For sufficiently large $i$, only members of
$T=(T_t)_{t\ge -L_T}\in \mathcal F$ with $L_T=\infty$
give rise to boxes of the form $T[-i,j]$
The result then follows.
just as in Case 2, with the comparison intervals
expanding to the left rather than to the right.
\end{proof}

\begin{theorem}\label{thm:affine-subshift}
Let $X\subseteq\Sigma^{\mathbb N}$ be a one-sided subshift over a
finite alphabet.  Suppose that $p_X(n)\le Cn$ and $r_X(n)\le D$ for
all sufficiently large $n$, where $C>0$ and $D\ge0$.  Then every
extendable finite box has dimension at most $C+D$.  For each
integer $d\ge1$, there are integers $a_d\ge0$ and $b_d$ such that
$$
B_X^{(d)}(N)=a_dN+b_d
$$
for all sufficiently large $N$.  Consequently, the total number of
positive-dimensional maximal extendable $N$-boxes is eventually
affine.
\end{theorem}

\begin{proof}
The dimension bound is Proposition~\ref{prop:affine-dim-bound}.
Let $M,\alpha_d$, and $h_d$ be as in
Lemma~\ref{lem:affine-stabilization-short}.  Every
positive-dimensional maximal extendable box has a unique first
branching position, and hence
$$
B_X^{(d)}(N)=\sum_{i=0}^{N-1}h_d(i,N-1-i).
$$
For $N\ge2M+1$, splitting off the two ends gives
$$
\begin{aligned}
B_X^{(d)}(N)={}&(N-2M)\alpha_d\\
&+\sum_{i=0}^{M-1}h_d(i,N-1-i)\\
&+\sum_{j=0}^{M-1}h_d(N-1-j,j).
\end{aligned}
$$
Each term in the last two sums is eventually constant by
Lemma~\ref{lem:affine-stabilization-short}.  Thus there is an integer $\beta_d$ such that $B_X^{(d)}(N)=\alpha_d N+\beta_d$ for all $N$ large.  By the dimension bound, only
finitely many positive dimensions can occur, so their total is
eventually affine as well.
\end{proof}

We finish by recovering the statement for individual infinite words.

\begin{proof}[Proof of Theorem \ref{thm:affine}]
Apply Theorem~\ref{thm:affine-subshift} to $X_{\bf w}$.
As observed at the beginning of the section,
$p_{X_{\bf w}}(n)=p_{\bf w}(n)$ and
$r_{X_{\bf w}}(n)\le1$ for every $n$.
\end{proof}

\section{Regularity for automatic words}
\label{sec:regular-modular}

In this section we consider the box counting functions
associated to $k$-automatic words. We show that if
${\bf w}$ is a $k$-automatic word, then the number
$B_{\bf w}(N)$ of maximal extendable $N$-boxes
is a $k$-regular sequence. Our approach uses
first-order logic and follows the general method
developed by Charlier, Rampersad, and Shallit
\cite{CRS}. We assume some familiarity with
first-order logic in this section.

Let ${\bf w}=w_0w_1\cdots$ be a $k$-automatic word
over a finite alphabet $\Sigma$, where $k\ge 2$.
For each $a\in\Sigma$, we associate to ${\bf w}$
a unary predicate $P_a$ on $\mathbb N$ by setting
$$
P_a(i)\iff w_i=a.
$$
We work with the first-order structure
$$
(\mathbb N,+,<,(P_a)_{a\in\Sigma}).
$$
Recall that a subset of $\mathbb N^r$ is
$k$-recognizable if the base-$k$ representations
of its elements, read as an $r$-tuple of digits and allowing padding with leading zeros to ensure a common
length, form a regular language. Since ${\bf w}$ is $k$-automatic,
each predicate $P_a$ defines a $k$-recognizable
set. Addition and order are also $k$-recognizable,
and the class of $k$-recognizable relations is
closed under the operations corresponding to
first-order quantifiers and logical connectives.
Consequently, every relation first-order
definable in the above structure is
$k$-recognizable.

We will also use the following counting result
from the theory of automatic sequences
(see \cite{CRS} for further details). Suppose that
$E\subseteq\mathbb N^{1+s}$ is a $k$-recognizable
relation, where $s$ is fixed, and that for
every $N$ the set
$$
E_N=\{{\bf x}\in\mathbb N^s:(N,{\bf x})\in E\}
$$
is finite. Then the sequence
$$
N\longmapsto |E_N|
$$
is $k$-regular. 

We now apply this approach to the enumeration of boxes. Recall
that every $k$-automatic word has linear
subword complexity. Thus, by
Lemma~\ref{lem:finite-right-patterns}, there
is an integer $H$ such that if an extendable
finite box has its first branching position
at $i$, then all its other branching positions
are at most $i+H$. 

In particular, every extendable finite box
has at most $R:=H+1$ branching positions.
This observation will allow us to describe
such boxes using a bounded number of integer
variables.  

We now develop some notation and terminology that will be useful in the enumeration results below.
We first explain how to encode boxes in terms of tuples of natural number.
Fix an integer $0\le r\le R$ and
non-singleton subsets
$M_1,\ldots,M_r\subseteq\Sigma$.
We call the tuple
${\bf M}=(M_1,\ldots,M_r)$ the \emph{type} of
a box whose branching positions, in
increasing order, have coordinate sets
$M_1,\ldots,M_r$. Notice that since
$R$ and $\Sigma$ are fixed, there are
only finitely many possible types.

We can now encode a box of type ${\bf M}$ by
a tuple
$$
(N,i,h_1,\ldots,h_r),
\qquad 0\le h_1<\cdots<h_r<N,
$$
which we call the \emph{code} of the box.
Here $N\ge1$ is the number of positions in our box, $h_1,\ldots ,h_r$ represent the branching positions, and 
$M_t$ is the coordinate set at position $h_t$.  The significance of $i$ is that since the other coordinates are size $1$ and since each word in a box is a subword of ${\bf w}$, we can choose $i$ so that the coordinate set of our box at each position $j\not\in \{h_1,\ldots ,h_r\}$ is given by 
$\{w_{i+j}\}$. 
We choose $i$ having the additional property
that
$$
w_{i+h_t}=\min M_t
~~~{\rm for}~t\in \{1,\ldots ,r\}.
$$
Thus the length-$N$ subword beginning
at position $i$ is the lexicographically
least word in the box. 

We also require
that $i$ is the \emph{first} position at which
this subword occurs in ${\bf w}$.
For a box all of whose elements occur in $\mathbf w$,
its lexicographically least word occurs at some point, so the
first-occurrence position $i$ exists and is unique.
These requirements ensure that every such box of
the given type has a unique code. Since there are
only finitely many length-$N$ subwords and possible
branching positions, we also have that the fibres of the projection map $(N,i,h_1,\ldots ,h_r)\mapsto N$ are finite.

We now work towards showing that we can define the maximal extendable boxes.  We do this in a sequence of steps, where we first use this information to show that the function sending $N$ to the number of $N$-boxes whose words are subwords of ${\bf w}$ is definable in this structure.

\begin{lemma}\label{lem:regular-factor-boxes}
Let ${\bf w}$ be a $k$-automatic right-infinite
word over a finite ordered alphabet $\Sigma$,
where $k\ge 2$, and let $R$ be a nonnegative
integer. Then the number of $N$-boxes $B$ having
at most $R$ branching positions and with
all words in $B$ subwords of ${\bf w}$ is
a $k$-regular function of $N$.
\end{lemma}

\begin{proof}
We let $B$ be an $N$-box of type ${\bf M}=(M_1,\ldots,M_r)$ and we let $(N,i,h_1,\ldots ,h_r)$ denote its corresponding code.
We show that the conditions defining
these codes are first-order. For
nonnegative integers $x,y$, let
$$
\operatorname{Same}(x,y)
\iff
\bigvee_{a\in\Sigma}
(P_a(x)\wedge P_a(y)).
$$
This formula expresses the condition
$w_x=w_y$. We can therefore express
equality of the length-$N$ subwords
beginning at positions $i$ and $j$
by the formula
$$
\operatorname{Eq}(i,j,N)
\iff
(\forall u<N)\operatorname{Same}(i+u,j+u).
$$
In particular, the condition that
$i$ is the first occurrence of its
length-$N$ subword is
$$
(\forall j<i)\neg\operatorname{Eq}(i,j,N).
$$
The remaining conditions defining a
code are also first-order, since
the sets $M_t$ are fixed.

It remains to express the condition
that every element of the box of type ${\bf M}$ encoded by $(N,i,h_1,\ldots ,h_r)$ is a
length-$N$ subword of ${\bf w}$.
For a fixed type ${\bf M}$, let
$$
\mathcal A_{\bf M}
=M_1\times\cdots\times M_r.
$$
An element
${\bf a}=(a_1,\ldots,a_r)$ of
$\mathcal A_{\bf M}$ determines a
word in the box encoded by $(N,i,h_1,\ldots ,h_r)$ by choosing $a_t$ at
position $h_t$ and keeping the
singleton coordinates unchanged.

For each such ${\bf a}$, let
$V_{\bf a}(j)$ denote the formula
$$
\begin{aligned}
V_{\bf a}(j)\iff{}&
\bigwedge_{t=1}^r P_{a_t}(j+h_t)\\
&\wedge(\forall u<N)
\left(
\left(\bigwedge_{t=1}^r u\ne h_t\right)
\Longrightarrow
\operatorname{Same}(i+u,j+u)
\right).
\end{aligned}
$$
Here $N,i,h_1,\ldots,h_r$ are understood
to be the parameters of the code.
The formula $V_{\bf a}(j)$ says that
the word determined by ${\bf a}$
occurs in ${\bf w}$ beginning at
position $j$.

Consequently, the condition that
every word in our box is a subword of
${\bf w}$ is expressed by
$$
\bigwedge_{{\bf a}\in\mathcal A_{\bf M}}
(\exists j)\,V_{\bf a}(j).
$$
Notice that this is a first-order
formula, since $\mathcal A_{\bf M}$
is a fixed finite set. 

We have now shown that for each
fixed type ${\bf M}$, the codes for boxes whose words are all subwords of ${\bf w}$
form a first-order definable relation
in
$$
(\mathbb N,+,<,(P_a)_{a\in\Sigma}),
$$
and hence a $k$-recognizable relation.
Moreover, for each fixed $N$ there
are only finitely many valid codes,
since there are only finitely many
length-$N$ subwords and finitely many
choices of branching positions and
coordinate sets.

By the results of \cite{CRS}, the number of boxes of type
${\bf M}$ is therefore a $k$-regular
function of $N$. Summing over the
finitely many possible types gives
the result.
\end{proof}

We next show that if one imposes the additional constraint that the boxes are extendable with respect to ${\bf w}$ then we can still define this collection of objects in our structure.

\begin{lemma}\label{lem:regular-extendable}
Let ${\bf w}$ be a $k$-automatic right-infinite word
over a finite ordered alphabet $\Sigma$, where $k\ge 2$,
and let $R$ be a nonnegative integer such that every
extendable finite box has at most $R$ branching positions.
Then the codes
representing extendable boxes form a $k$-recognizable
relation. Consequently, the number of extendable
$N$-boxes is a $k$-regular function of $N$.
\end{lemma}

\begin{proof}
Fix a type ${\bf M}=(M_1,\ldots,M_r)$ with $r\le R$,
and recall that
$$
\mathcal A_{\bf M}=M_1\times\cdots\times M_r
$$
is the set of choices of letters at the branching
positions. We use the notation from the proof of
Lemma~\ref{lem:regular-factor-boxes}. In particular,
for each ${\bf a}\in\mathcal A_{\bf M}$, the formula
$V_{\bf a}(j)$ expresses the condition that the
word corresponding to ${\bf a}$ in a box of type ${\bf M}$ occurs in
${\bf w}$ beginning at position $j$.

We wish to express the condition that all words 
of a box of type ${\bf M}$ admit a common infinite right continuation.
Notice that by K\"onig's infinity lemma this holds if and only if, for every
nonnegative integer $m$, there is a word of length
$m$ that is a common right continuation of all the
vertices. We can express this condition using
first-order logic.

To see this, fix ${\bf a}^0\in\mathcal A_{\bf M}$.
For each ${\bf a}\in\mathcal A_{\bf M}$, we introduce
a variable $j_{\bf a}$ giving the position of an
occurrence of the word in our box corresponding to ${\bf a}$. We then
consider the formula
$$
\begin{aligned}
(\forall m)(\exists (j_{\bf a})_{
{\bf a}\in\mathcal A_{\bf M}})\quad
&\bigwedge_{{\bf a}\in\mathcal A_{\bf M}}
V_{\bf a}(j_{\bf a})\\
&\wedge(\forall u<m)
\bigwedge_{{\bf a}\in\mathcal A_{\bf M}}
\operatorname{Same}(j_{\bf a}+N+u,
j_{{\bf a}^0}+N+u).
\end{aligned}
$$
The first part of this formula says that every
word occurs in ${\bf w}$ at its corresponding
position $j_{\bf a}$. The second part says that
the length-$m$ subwords immediately following
these occurrences are all equal.

Thus the formula expresses the condition that
the words in a box admit common right continuations
of arbitrarily large finite length. As we noted earlier,
this is equivalent to the existence of a common
infinite right continuation. Thus the formula above holds precisely when
a box of type ${\bf M}$ is extendable.

Finally, notice that $\mathcal A_{\bf M}$ is a
fixed finite set, since the type ${\bf M}$ is
fixed. In particular, the formula above involves
only finitely many variables, independently of
$N$. It follows that the codes representing
extendable boxes of type ${\bf M}$ form a
first-order definable relation, and hence a
$k$-recognizable relation.

As we saw earlier, these
codes have finite fibres over $N$ and so by \cite{CRS} the number of extendable boxes
of type ${\bf M}$ is a $k$-regular function
of $N$. Since there are only finitely many
possible types and every extendable box has
at most $R$ branching positions, summing over
all types gives the result.
\end{proof}

Now to finish we just have to show that maximality of extendable boxes is first-order definable.

\begin{lemma}\label{lem:regular-maximal}
Let ${\bf w}$ be a $k$-automatic right-infinite word
over a finite ordered alphabet $\Sigma$, where $k\ge 2$,
and let $R$ be a nonnegative integer such that every
extendable finite box has at most $R$ branching positions.
Then, for each type ${\bf M}=(M_1,\ldots M_r)\subseteq \Sigma^r$ with $r\le R$, the codes
representing maximal extendable boxes of type ${\bf M}$ form a
$k$-recognizable relation.
\end{lemma}

\begin{proof}
Let $S=S_0\times\cdots\times S_{N-1}$ be the box
represented by a code of type ${\bf M}$.
For $0\le h<N$ and $a\in\Sigma\setminus S_h$,
let $S^{+(h,a)}$ denote the box obtained from $S$
by replacing $S_h$ with $S_h\cup\{a\}$ and
leaving all other coordinates unchanged.

Notice that $S$ is maximal extendable if and only
if it is extendable and none of the boxes
$S^{+(h,a)}$ is extendable. 

We claim that this condition is definable in our structure. First, the condition
$a\in S_h$ is first-order definable from the
code. To see this, recall that a code of type
${\bf M}=(M_1,\ldots,M_r)$ has the form
$(N,i,h_1,\ldots,h_r)$. If $h=h_t$ for some $t$,
then $a\in S_h$ is equivalent to $a\in M_t$.
Otherwise, it is equivalent to $P_a(i+h)$.
Since the sets $M_t$ are fixed, this gives a
first-order formula.

We next consider extendability of $S^{+(h,a)}$.
There are two possibilities. If $h=h_t$ for
some $t$, then the enlarged box is obtained
by replacing $M_t$ with $M_t\cup\{a\}$.
Otherwise, we introduce one additional
branching position at $h$, with coordinate
set $\{w_{i+h},a\}$.

In either case, the enlarged box has at most
$R+1$ branching positions. We can therefore
use the common-continuation formula from the
proof of Lemma~\ref{lem:regular-extendable}
to express its extendability. Notice that
we do not require the description of the
enlarged box to be canonical. We only need
to describe its vertices using the original
code together with $h$ and $a$. Since
$\Sigma$ and $R$ are fixed, this involves
only finitely many variables and their number does not depend on $N$. Thus the condition that
$S^{+(h,a)}$ is extendable is first-order
definable.

Writing $\operatorname{Ext}(S)$ for the
first-order condition that $S$ is extendable,
we have shown that maximality is expressed by
$$
\operatorname{Ext}(S)\ \wedge\
(\forall h<N)\bigwedge_{a\in\Sigma}
\left(
a\in S_h\ \vee\
\neg\operatorname{Ext}(S^{+(h,a)})
\right).
$$
Consequently, the codes representing maximal
extendable boxes of type ${\bf M}$ form a
first-order definable relation in
$$
(\mathbb N,+,<,(P_a)_{a\in\Sigma}),
$$
and hence a $k$-recognizable relation.
\end{proof}

\begin{theorem}\label{thm:regular-modular}
Let ${\bf w}$ be a $k$-automatic right-infinite
word over a finite alphabet $\Sigma$, where $k\ge 2$.
Then the sequence $B_{\bf w}(n)$ counting maximal
extendable $n$-boxes is $k$-regular.
\end{theorem}

\begin{proof}
Recall that every $k$-automatic word has linear
subword complexity. By
Lemma~\ref{lem:finite-right-patterns}, there is
an integer $R$ such that every extendable
finite box has at most $R$ branching positions.

By Lemma~\ref{lem:regular-factor-boxes}, these
extendable boxes admit unique codes belonging to one of
finitely many types. Moreover, for each fixed
type, the codes form a $k$-recognizable relation
with finite fibres over $N$.
By Lemmas~\ref{lem:regular-extendable} and
\ref{lem:regular-maximal}, the condition that
such a code represents a maximal extendable
box is first-order definable, and hence
the corresponding codes also form a
$k$-recognizable relation.

The finite-fibre counting principle now shows
that the number of maximal extendable $n$-boxes
of each fixed type is a $k$-regular function
of $n$. Since there are only finitely many
types, their sum $B_{\bf w}(n)$ is also $k$-regular.
\end{proof}

We finish with a few examples illustrating the
possible behaviour of these box counting functions for automatic words.
\begin{exam} The periodic word
${\bf w}=(01)^\infty$ has $B_{\bf w}(n)=2$ for $n\ge 1$.
\end{exam}
Note that the orbit closure of ${\bf w}$
consists of two points, and no two distinct
points have a common right tail. Thus every
maximal extendable $N$-box is a singleton and there are two of them. Hence
$$
B_{\bf w}^{(0)}(N)=2,\qquad
B_{\bf w}^{(d)}(N)=0\quad {\rm for}~d\ge1.
$$

In contrast, if ${\bf w}=10^\infty$.
Its length-$N$ subwords are $0^N$ and
$10^{N-1}$, which admit the common right
continuation $0^\infty$. Consequently,
it has a unique maximal extendable $N$-box given by
$$
\{0,1\}\{0\}^{N-1},
$$
and so in this case
$$
B_{\bf w}^{(1)}(N)=1,\qquad
B_{\bf w}^{(d)}(N)=0\quad(d\ne1).
$$
\begin{exam}
If ${\bf w}=w_0w_1\cdots$ is the word, where
$$
w_n=
\begin{cases}
1 & \text{if }n\text{ is a power of }2,\\
0 & \text{otherwise}.
\end{cases}
$$
then using Walnut we find
$$
B_{\bf w}^{(0)}(N)=p_{\bf w}(N)-N-1, B_{\bf w}^{(1)}(N)=N,~{\rm and}~
B_{\bf w}^{(d)}(N)=0\quad {\rm for}~d\ge2.
$$
\end{exam}
We note that Walnut software allows us to compute $p_{\bf w}(n)$ and we see that for $m\ge2$,
$$
p_{\bf w}(2^m)=3\cdot 2^{m-1}+2,\qquad
p_{\bf w}(3\cdot2^{m-1})=5\cdot2^{m-1}+2.
$$
Thus $B_{\bf w}^{(0)}(N)$ is not
eventually affine, even though ${\bf w}$
has linearly bounded subword complexity.
\section{Sharpness of the linear-complexity hypothesis}
\label{sec:sharpness}

Theorem \ref{thm:affine} shows that for words of linear subword complexity,
the number of maximal extendable boxes of any fixed
positive dimension is eventually affine. We now show
that this conclusion need not hold if the complexity
is allowed to grow even arbitrarily slowly above linear.

Our construction is similar in spirit to constructions
of Greenfeld, Moreira, and Zelmanov \cite{GMZ} and
Greenfeld, Leemann, Moreira, and Zelmanov \cite{GLMZ},
who use sparse sets of positions and independently
variable letters to study subword complexity and its
growth. The additional difficulty in our setting is
to ensure that the boxes we construct are maximal.
We accomplish this by introducing three marker letters
that allow us to recover the type and position of a
pattern even after one of its letters has been changed.

\begin{theorem}\label{thm:near-linear-sharpness}
Let $f:\mathbb N\to\mathbb R_{>0}$ be a weakly
increasing function such that $f(N)\to\infty$.
Then there is a right-infinite word ${\bf w}$ over
an alphabet of five letters such that
$$
p_{\bf w}(N)=O(Nf(N))
$$
and, for every positive integer $d$,
$$
\lim_{N\to\infty}\frac{B_{\bf w}^{(d)}(N)}{N}
=\infty.
$$
\end{theorem}

We construct the word as follows. Let
$\Sigma=\{0,1,A,B,C\}$ and choose a sequence of
positive integers $d_1,d_2,\ldots$ in which every
positive integer occurs infinitely often.
Since $f(N)\to\infty$, we can choose a strictly
increasing sequence of positive integers
$g_1,g_2,\ldots$ such that
$$
f(g_j)\ge \sum_{i=1}^j(d_i+3)2^{d_i}
$$
for every $j\ge 1$.

For each $j$ and
$\varepsilon=(\varepsilon_1,\ldots,\varepsilon_{d_j})
\in\{0,1\}^{d_j}$, we define a word
$U_{j,\varepsilon}$ of length
\begin{equation}
L_j=(d_j+2)g_j+1.\end{equation}
The letters at positions
$0,g_j,2g_j$ are $A,B,C$, respectively,
and the letter at position $(r+2)g_j$
is $\varepsilon_r$ for $1\le r\le d_j$.
All other letters are zero.

Thus each $j$ gives $2^{d_j}$ words of length
$L_j$. We will refer to these as the words
of type $j$.

We now choose a sequence $V_1,V_2,\ldots$
of these words such that every
$U_{j,\varepsilon}$ occurs infinitely often.
We construct ${\bf w}$ by placing $V_t$
at position $b_t$, where $b_1=0$ and
the positive integers $b_t$ are chosen
inductively so that
$$
b_{t+1}\ge 2(b_t+|V_t|).
$$
All positions not occupied by one of
the words $V_t$ are filled with zeros.  We call the occurrences of $V_t$ in ${\bf w}$ the \emph{blocks} of ${\bf w}$ and a block of the form $U_{j,\epsilon}$ will be called a \emph{block of type} $j$.
In addition we define 
\begin{equation} E_t=b_t+|V_t|
\end{equation}
so our condition above gives
$
b_{t+1}\ge 2E_t$.

Notice that the gaps between successive blocks
tend to infinity, and so there are increasingly long strings of zeros between consecutive blocks. In particular each fixed
word $U_{j,\varepsilon}$ occurs infinitely
often with arbitrarily long strings of zeros
on both sides.

\begin{lemma}\label{lem:sharp-complexity}
Let $f$ and ${\bf w}$ be as in the preceding
construction. Then for every $N\ge 1$ we have
$$
p_{\bf w}(N)\le 5N+1+Nf(N).
$$
\end{lemma}

\begin{proof}
We count the length-$N$ subwords of ${\bf w}$
according to how many of the blocks $V_t$
contribute nonzero letters.

Suppose first that a length-$N$ subword
contains nonzero letters from at least two
different blocks. Let $s$ be its starting
position, and let $V_t$ be the first block
contributing a nonzero letter. Then $s<E_t$,
and since the subword also contains a nonzero
letter from a later block, we have
$$
s+N>b_{t+1}\ge 2E_t.
$$
Consequently $N>E_t>s$. Thus every such
subword begins at a position $s<N$, and
there are at most $N$ possibilities.

Now consider a length-$N$ subword whose
nonzero letters all come from a single
block of type $j$. If $g_j\ge N$, then
the nonzero letters of this block are
separated by at least $N$ positions.
Thus our subword contains at most one
nonzero letter. There are at most $4N+1$
words of length $N$ having at most one
nonzero letter, since we can choose its
position and one of the four nonzero
letters, or take the word $0^N$.

It remains to consider blocks of type
$j$ with $g_j<N$. Fix one of the
$2^{d_j}$ words $U_{j,\varepsilon}$.
A length-$N$ subword containing a nonzero
letter from this block must contain at
least one of the $d_j+3$ distinguished
positions. There are at most $N$ ways
to place any one of these positions
inside the subword. Consequently, the
number of length-$N$ subwords arising
from blocks of type $j$ is at most
$$
N(d_j+3)2^{d_j}.
$$

Now let
$J$ denote the largest such integer for which $g_J<N$. By our
choice of the $g_j$ and the monotonicity
of $f$, we have
$$
\sum_{g_j<N}(d_j+3)2^{d_j}
\le f(g_J)\le f(N).
$$

Combining the three estimates gives
$$
p_{\bf w}(N)\le N+(4N+1)+Nf(N),
$$
as required.
\end{proof}
We prove a lemma that allows us to show certain boxes are maximal.

\begin{lemma}\label{lem:sharp-orbit-closure}
Suppose that ${\bf x}\in X_{\bf w}$ contains at least
two distinct letters from $\{A,B,C\}$. Then
either ${\bf x}$ is a shift of ${\bf w}$, or
${\bf x}$ is a right-infinite suffix of
$$
0^m U_{j,\varepsilon}0^\infty
$$
for some $m\ge0$ and some $j\ge1$, and
$\varepsilon\in\{0,1\}^{d_j}$.
\end{lemma}

\begin{proof}
Suppose ${\bf x}$ is not a shift of ${\bf w}$.
We can write
$$
{\bf x}=\lim_{n\to\infty}\sigma^{m_n}({\bf w}),
$$
where $m_n\to\infty$.

Choose two positions in ${\bf x}$ occupied by
distinct letters from $\{A,B,C\}$. For all
sufficiently large $n$, the same letters
occur at these positions in
$\sigma^{m_n}({\bf w})$. Since the gaps
between successive blocks tend to infinity,
these markers must belong to the same
block $V_{t_n}$.

The letters that occur from $\{A,B,C\}$ and their distance determine
$g_j$, and hence $j$, since the markers
$A,B,C$ occur at positions $0,g_j,2g_j$
and the integers $g_j$ are distinct.
Their positions also determine where the block
occurs relative to ${\bf x}$.

There are only finitely many 
$U_{j,\varepsilon}$ of this type. Passing
to a subsequence, we may assume that the
$U_{j,\epsilon}$ is fixed. Since the gaps on
either side of $V_{t_n}$ tend to infinity,
the limit ${\bf x}$ is a right-infinite
suffix of this block padded by zeros.
\end{proof}

We now prove the remaining assertions of
Theorem~\ref{thm:near-linear-sharpness}.

\begin{proof}[Proof of Theorem~\ref{thm:near-linear-sharpness}]
The complexity estimate follows from
Lemma~\ref{lem:sharp-complexity}.
We fix a positive integer $d$ and show
that $B_{\bf w}^{(d)}(N)/N\to\infty$.

Suppose that $j$ is an index with $d_j=d$.
For $N\ge L_j$ and $0\le r\le N-L_j$,
we define an $N$-box $B_{j,r}(N)$ as
follows. At positions
$$
r,\quad r+g_j,\quad r+2g_j
$$
we take the singleton sets
$\{A\},\{B\},\{C\}$, respectively.
At the $d$ positions
$$
r+3g_j,\ r+4g_j,\ldots,r+(d+2)g_j
$$
we take the set $\{0,1\}$, and
at every other position we take
the singleton set $\{0\}$.

This box has dimension $d$. Moreover,
it is extendable, since every word in $B_{j,r}(N)$ is obtained by placing
some $U_{j,\varepsilon}$ inside a string
of zeros. As every realization of type
$j$ occurs infinitely often with
arbitrarily long zero gaps on both
sides, each word in the box has $0^\infty$
as a right continuation in $X_{\bf w}$.

We claim that these boxes are maximal
extendable. To see this, by Lemma \ref{lem:sharp-orbit-closure}
if ${\bf x}\in X_{\bf w}$
contains at least two distinct letters
from $\{A,B,C\}$ then if ${\bf x}$ is not
a shift of ${\bf w}$, then ${\bf x}$
is obtained by taking a word
$U_{j,\varepsilon}$, adjoining zeros
on either side, and taking a
right-infinite suffix.

Now suppose that $B_{j,r}(N)$ is
contained in a larger extendable box.
Since extendability is preserved
under passage to subboxes, there
must be some coordinate at which
we can adjoin an additional letter
while preserving extendability.

Let ${\bf z}$ be a common infinite
right continuation of such an
enlarged box. In particular, two
distinct vertices $u,v$ of
$B_{j,r}(N)$ satisfy
$$
u{\bf z},v{\bf z}\in X_{\bf w}.
$$
We claim that ${\bf z}=0^\infty$.

First suppose that ${\bf z}$ is not
eventually zero. Since $u$ and $v$
both contain the three markers
$A,B,C$, the preceding observation
shows that $u{\bf z}$ and $v{\bf z}$
must both be shifts of ${\bf w}$.
But ${\bf w}$ is not ultimately
periodic, since it contains
infinitely many occurrences of $A$
and arbitrarily long zero gaps.
Consequently, two distinct shifts
of ${\bf w}$ cannot have the same
right-infinite tail. This contradicts
the fact that $u\ne v$.

It follows that ${\bf z}$ is eventually
zero. Thus $u{\bf z}$ is eventually
zero and cannot be a shift of
${\bf w}$. By the observation above,
it must come from a single padded
block $U_{k,\varepsilon}$.
The positions of $A,B,C$ completely determine
the type of this
block, so $k=j$. Since the entire
block is contained in the first
$N$ positions of $u{\bf z}$, we
conclude that ${\bf z}=0^\infty$.

Now consider a word $v'$ obtained
by choosing an additional letter
at one coordinate of $B_{j,r}(N)$.
If the enlarged box were extendable,
we would have
$$
v'0^\infty\in X_{\bf w}.
$$
Notice that $v'$ still contains at
least two of the original markers
$A,B,C$, since only one coordinate
has been changed. These two markers
determine the type $j$ and the
position of the corresponding block.
It follows from our observation
that $v'0^\infty$ must arise from
the same type-$j$ pattern.

But this is impossible since every realization
of type $j$ has the letters $A,B,C$
at the three prescribed positions,
has letters in $\{0,1\}$ at the
$d$ variable positions, and has
zeros everywhere else. The word
$v'$ differs from this pattern at
one coordinate. Thus no single-letter
enlargement of $B_{j,r}(N)$ can
be extendable, proving that
$B_{j,r}(N)$ is maximal.

We have therefore constructed
$N-L_j+1$ distinct maximal extendable
$N$-boxes of dimension $d$ for
each index $j$ with $d_j=d$ and
$L_j\le N$. Moreover, the boxes
arising from different choices
of $j$ and $r$ are distinct, since
the positions of the three markers
determine both $g_j$ and $r$.

Consequently,
$$
B_{\bf w}^{(d)}(N)
\ge
\sum_{\substack{j:\ d_j=d\\L_j\le N}}
(N-L_j+1).
$$
Since every positive integer occurs
infinitely often in the sequence
$d_1,d_2,\ldots$, there are infinitely
many indices $j$ with $d_j=d$.
For any fixed positive integer $m$,
choose $m$ distinct such indices.
Dividing the preceding inequality
by $N$ and letting $N\to\infty$
gives
$$
\liminf_{N\to\infty}
\frac{B_{\bf w}^{(d)}(N)}N\ge m.
$$
Since $m$ is arbitrary, we conclude
that
$$
\lim_{N\to\infty}
\frac{B_{\bf w}^{(d)}(N)}N=\infty.
$$
The result follows.
\end{proof}

\section{An application to monomial algebras}
\label{sec:monomial-application}

In this brief section we apply the combinatorial results to give a description of point modules
of monomial algebras. Throughout, $\mathcal P_n(A)$
denotes the reduced projection of the space of genuine
infinite point modules onto its first $n+1$ projective
coordinates, as defined in
\S\ref{sec:algebraic-background}. We note that this is a more restrictive condition than taking the scheme of all truncated point modules.

\begin{theorem}\label{thm:monomial-algebra}
Let $K$ be an algebraically closed field and let
$A=K\langle x_1,\ldots,x_m\rangle/I$ be a finitely
generated semiprime monomial algebra satisfying
$$
\sum_{i=0}^n\dim_K A_i=O(n^2).
$$
Then for every integer $d\ge1$ there are integers
$a_d\ge0$ and $b_d$ such that, for all sufficiently
large $n$, the number of $d$-dimensional irreducible
components of $\mathcal P_n(A)$ is $a_dn+b_d$.
The total number of positive-dimensional irreducible
components is likewise eventually affine as a function of $n$.
\end{theorem}

\begin{proof}
Let $\Sigma=\{x_1,\ldots,x_m\}$ and let
$\mathcal L(A)$ be the language of nonzero monomials.
For a semiprime ring, $uAu=0$ implies $u=0$.
Thus if $u$ is a nonzero monomial, there is a monomial
$v$ such that $uvu$ is nonzero. As $I$ is monomial,
this implies that every word in $\mathcal L(A)$
admits a longer nonzero extension on each side.
Iterating, its extensions can be made arbitrarily
long. In particular, $\mathcal L(A)$ is factorial,
left-prolongable and right-prolongable.

Put $p(n)=\dim_K A_n=\#\mathcal L_n(A)$. Since
every word has a right extension, $p(n)$ is
nondecreasing. The growth hypothesis therefore
gives
$$
np(n)\le\sum_{i=n}^{2n}p(i)=O(n^2),
$$
so $p(n)=O(n)$.

Let $X\subseteq\Sigma^{\mathbb N}$ be the subshift
consisting of right-infinite words all of whose
finite subwords lie in $\mathcal L(A)$. Closure under the process of taking subwords together
with K\"onig's infinity lemma show that every
word in $\mathcal L(A)$ is a prefix of a point of
$X$. Thus $\mathcal L(X)=\mathcal L(A)$, and $X$
has linear subword complexity and a left-prolongable
language. Theorem~\ref{thm:affine-subshift} applies.

By \cite[Theorem 6.8]{BG}, the
irreducible components of $\mathcal P_n(A)$
correspond bijectively to maximal extendable
$(n+1)$-boxes for $X$, with the same dimensions.
Theorem~\ref{thm:affine-subshift} therefore gives
eventual affine counts in each positive dimension. The total positive-dimensional count
is eventually affine by the same theorem.
\end{proof}

\section{Questions}
\label{sec:questions}
The results of this paper give rather precise information
about positive-dimensional irreducible components, but
tell us considerably less about isolated components.
We finish with a few ring theoretic questions that I find interesting, and which I have no idea how to answer.  Maybe some person or machine out there will consider these problems and solve them.
\begin{question}
For a $k$-automatic word ${\bf w}$, is the number
of connected components of the variety
$\mathcal P_{N}(A_{\bf w})$ always $k$-regular?
\end{question}
The main subtlety in this question is that irreducible components of $\mathcal{P}_N(A_{\bf w})$ can intersect and so the number of connected components can be strictly smaller than the total number of irreducible components.  There is, however, a purely word theoretic way of posing this question as follows.
We make a graph $G_N$ whose vertices are the length-$(N+1)$ subwords of ${\bf w}$. We join words $u$ and $v$ with an edge if they differ in exactly one coordinate and there is an infinite word ${\bf z}$ with $u{\bf z},v{\bf z}\in X_{\mathbf w}$. Then the number of connected components of 
$\mathcal P_{N}(A_{\mathbf w})$ is just the number of connected components of the graph $G_N$
Indeed, the words in each extendable box are connected by its coordinate edges, and two irreducible components of $\mathcal{P}_{N}(A_{\bf w})$ intersect exactly when their boxes share a common word. I thought about trying to apply the first-order methods of \cite{CRS} here, but it would require some additional information about the depth of the graph $G_N$, since connectedness is a second-order property in general and not first-order.

A final interesting question is whether Theorem \ref{thm:monomial-algebra} extends to arbitrary semiprime graded algebras of quadratic growth.
For a general connected graded algebra
$A=\bigoplus_{n\ge0}A_n$ generated in degree
one, we let $\mathcal P_N(A)$ denote the
reduced variety obtained by retaining the first
$N+1$ projective coordinates of genuine infinite
point modules. Equivalently, writing
$A_{\le N+1}=\bigoplus_{i=0}^{N+1}A_i$,
we may regard $\mathcal P_N(A)$ as the reduced
subvariety of the Grassmannian variety
$$
\operatorname{Gr}_{\dim_K A_{\le N+1}-(N+2)}
(A_{\le N+1})
$$
corresponding to the degree-$(N+1)$
truncations of genuine infinite point modules.

\begin{question} Let $A=\bigoplus A_i$ be a finitely generated connected graded semiprime algebra over an algebraically closed base field $K$ that is generated in degree one.  If $A$ has quadratic growth, that is, if $\sum_{i\le n} {\rm dim}_K(A_i) = O(n^2)$, for a fixed $d\ge 1$ is the number of $d$-dimensional irreducible components of $\mathcal{P}_N(A)$ eventually affine in $N$?\label{q:2}
\end{question}
We note that there are very strong structural results in the
noetherian setting. Artin and Stafford
\cite{AS95} showed that a connected graded
domain of Gelfand--Kirillov dimension two,
generated in degree one, agrees in sufficiently
large degrees with a twisted homogeneous
coordinate ring of a projective curve. They
subsequently gave a geometric description of
noetherian semiprime graded algebras of
quadratic growth \cite{AS00}. Moreover, by
Artin and Zhang \cite[Corollary E4.12]{AZ01},
the schemes parametrizing truncated point
modules of a strongly noetherian algebra
generated in degree one eventually stabilize.
Consequently, the component counts in the
preceding question are eventually constant
in the strongly noetherian case. It would
be interesting to understand how much of
this rigidity persists when one removes the noetherian hypothesis.

We note that the examples given in \S\ref{sec:monomial-application} show that if one works instead with Gelfand-Kirillov dimension two rather than quadratic growth then Question 
\ref{q:2} has a negative answer in general. 
\section*{AI Declaration}
What a bizarre world we live in!  The impetus for this project came from my overhearing Bon Jovi's confessional ballad \emph{Bed of Roses} while running errands. As the song played, I found myself increasingly enraged by the singer's implacable desire to lay a woman down on a bed of roses\footnote{In the song, this urge acquires a particularly grim urgency from the fact that he himself sleeps on a bed of nails.} and I began thinking about the box-counting functions considered in this paper.  I realized that for automatic words the problem of proving $k$-regularity could be approached using techniques from first-order logic (specifically from the paper of Charlier, Rampersad, and Shallit \cite{CRS}). I rushed home and discussed the idea with ChatGPT, which proceeded to give the statement of Theorem \ref{thm:affine} about words of linear subword complexity. I accordingly disclose that generative AI played a substantial role in the development of the mathematical arguments, as well as in the preparation and revision of the manuscript itself. I find myself less certain how to acknowledge the contribution of Bon Jovi.

\end{document}